\def\NCS{N_{\rm CS}}

\def\gsim{\mbox{~{\raisebox{0.4ex}{$>$}}\hspace{-1.1em}
	{\raisebox{-0.6ex}{$\sim$}}~}}
\def\lsim{\mbox{~{\raisebox{0.4ex}{$<$}}\hspace{-1.1em}
	{\raisebox{-0.6ex}{$\sim$}}~}}

\newfam\scrfam                                          
\global\font\twelvescr=rsfs10 scaled\magstep1%
\global\font\eightscr=rsfs7 scaled\magstep1%
\global\font\sixscr=rsfs5 scaled\magstep1%
\skewchar\twelvescr='177\skewchar\eightscr='177\skewchar\sixscr='177%
\textfont\scrfam=\twelvescr\scriptfont\scrfam=\eightscr
\scriptscriptfont\scrfam=\sixscr

\documentstyle[preprint,aps,fixes,epsf,eqsecnum]{revtex}

\makeatletter           
\@floatstrue
\def\figure{\let\@capwidth\columnwidth\@float{figure}}
\let\endfigure\end@float
\@namedef{figure*}{\let\@capwidth\textwidth\@dblfloat{figure}}
\@namedef{endfigure*}{\end@dblfloat}
\def\table{\let\@capwidth\columnwidth\@float{table}}
\let\endtable\end@float
\@namedef{table*}{\let\@capwidth\textwidth\@dblfloat{table}}
\@namedef{endtable*}{\end@dblfloat}
\def\la{\label}
\makeatother

\tightenlines
\newcommand\pcite[1]{\protect{\cite{#1}}}

\begin {document}


\preprint {UW/PT 01--23}

\title
    {
	Problems with lattice methods for electroweak preheating
    }

\author{Guy D. Moore}
\address
    {%
    Department of Physics,
    University of Washington,
    Seattle, Washington 98195
    }%

\date {February 2001}

\maketitle
\vskip -20pt

\begin {abstract}%
    {%
Recently Garcia Bellido et.\ al.\ have proposed that electroweak
baryogenesis may occur at the end of inflation, in a scenario where the
reheat temperature is too low for electroweak symmetry 
restoration.  I show why the scenario is difficult to
test reliably by classical field techniques on the lattice.
    }%
\end {abstract}

\thispagestyle{empty}

\section {Introduction}
\la{sec:intro}

Inflation is a plausible and popular explanation for how the initial
cosmological epoch produces a universe with such startling size,
flatness, and homogeneity \cite{Guth}.  However, inflation makes even
more puzzling the other remarkable feature of the universe--that it
contains a macroscopic but relatively small net abundance of baryonic
matter (approximately 5 baryons per $10^{10}$ photons \cite{Tytler}).

Recently Garcia-Bellido, Grigoriev, Kusenko, and Shaposhnikov have
proposed a way that certain inflationary scenarios may be able to explain
baryogenesis (the origin of the baryon abundance) as well \cite{GB}.  During
inflation, most of the energy density in the universe is in the
potential energy of a scalar field, the inflaton.  It has recently been
understood that inflation can end, and the energy density stored in the
inflaton field can be converted into a thermal bath, much more abruptly
than had previously been thought possible, a process called
``preheating'' \cite{param_resonance}.  In the scenario of Garcia-Bellido
et.al., the baryons are created during this far from equilibrium
process.  

Baryon number is not conserved in the standard model \cite{tHooft}.  The
violation arises from nonperturbative physics of the SU(2) weak gauge
fields (W and Z bosons).  When occupation numbers of infrared fields
become large, nonperturbative physics can become efficient.  This
happens at high temperature and can also happen in other high excitation
situations.
In equilibrium, it occurs when there is no Higgs field condensate,%
\footnote
    {%
	The notion of a Higgs field condensate is a perturbative one,
	and should be used with some caution.  Nonperturbatively 
	speaking there is no qualitative distinction between symmetry
	broken and restored phases \pcite{Fradkin} and there can either 
	be a phase transition or an analytic crossover between them,
	depending on coupling parameters \pcite{KLRS_cross}.%
    } 
which requires a temperature 
$T \gsim 100$GeV.  At lower temperatures there is a Higgs field
condensate and baryon number violation is exponentially slow
\cite{ArnoldMcLerran}.  In the Garcia-Bellido scenario, the energy
density in the inflaton passes first into very infrared field modes, and
baryon number is readily violated; but when the fields fully thermalize
the temperature is low enough that there is a Higgs field condensate and
no further baryon number violation occurs.

The Garcia-Bellido scenario involves nonlinear, nonperturbative physics,
and its quantitative study is difficult to conduct analytically.  For
preheating in general, classical field techniques have proven useful
\cite{Khlebnikov,Toms}.  Baryon number violation in equilibrium has also
been treated accurately by classical field techniques
\cite{Ambjorn,Ambjorn2,slavepaper,Ambjorn3,others}, and recently
these techniques have been applied to the Garcia-Bellido scenario, both
in a 1+1 dimensional toy model \cite{GB,GB2} and in a more realistic 3+1
dimensional setting \cite{Rajantie}.  

It sounds natural to expect classical field techniques to work well
for the study of the Garcia-Bellido scenario.  This note will argue that
there are serious complications, because in the context of 3+1
dimensional Yang-Mills theory at realistic coupling, the techniques as
they exist to date contain spurious physics which can lead to ``fake''
early thermalization and baryon number violation.  At best, classical
field techniques will probably have to be modified and used with care in
this context; at worst they may not be useful at all.

\section{Classical Yang-Mills and the Lattice}

\subsection{Classical field approach}

First I summarize how classical field techniques are applied to
preheating.  The early stage of preheating {\em can} be understood
analytically.  When the inflaton field condensate begins to oscillate
about its potential minimum, certain field modes coupled to it have
their field amplitudes grow exponentially due to parametric resonance
\cite{param_resonance}.  In \cite{Khlebnikov,Toms} it is shown that
classical fields, initially populated with ``quantum vacuum initial
conditions,'' give quantitatively the same behavior.  Here ``quantum
vacuum initial conditions'' means Gaussian random initial conditions in
$k$ space, with mean squared excitation equal to the vacuum zero-point
excitations of the relevant modes, ($\hbar=1$)
\begin{equation}
\langle Q^2(k,t_{\rm init}) \rangle_{\rm q. \; vac} 
	= \frac{1}{2\sqrt{k^2+m^2}} \, , \qquad
\langle P^2(k,t_{\rm init}) \rangle_{\rm q. \; vac} 
	= \frac{\sqrt{k^2+m^2}}{2} \, , 
\label{eq:vac_init}
\end{equation}
with $Q$ and $P$ real canonical field and momentum variables.  When
parametric amplification has progressed the physics becomes nonlinear;
but the occupation numbers of the relevant field modes are large by
then, and the classical field approximation is valid.  The classical
field theory can be discretized on a lattice and evolved in real time by
standard algorithms.  This takes fully into account the nonlinear
interactions of different field modes.  Provided all of the interesting
physics involves modes well to the infrared of the lattice spacing
scale, the discretization should not disturb the physics of interest.

Such ``quantum'' initial conditions will not be preserved by
classical field evolution; nothing prevents the energy associated with
the ``zero point'' fluctuations, which have been turned into excitations
of classical fields, from moving between Fourier components of the field. 
They will not do so, or will do so very slowly, if the coupling constant
is small.  But at finite coupling, even without
an oscillating scalar (inflaton) condensate, this is not so.  Instead,
the lattice, classical field evolution will eventually approach the
classical thermal distribution for the lattice system in question, which
at weak coupling has%
\footnote
    {There are corrections to 
    $\langle Q^2 \rangle$ of relative size 
    $\lambda T/ \sqrt{k^2 + m^2}$, due to interactions.  However,
    if $P$ enters the Hamiltonian only quadratically and is not
    involved in constraints (such as Gauss' Law for $E$ in gauge
    theories) then the expression shown is exact and defines $T$.
    }
\begin{equation}
\langle Q^2(k) \rangle_{\rm thermal \; latt} 
	\simeq \frac{T}{\tilde{k}^2 + m^2} 
	\, , \qquad
\langle P^2(k) \rangle_{\rm thermal \; latt}
	= T \, ,
\label{eq:therm}
\end{equation}
for some $T \sim 1/a$.
(By $\tilde{k}^2$ I mean the
lattice dispersion relation, $a^2 \tilde{k}^2=4\sum_i \sin^2(k_i a/2)$.)  
Note that the distribution is {\em not} Lorentz invariant--Lorentz
invariance is broken by the lattice discretization--and has {\em
smaller} field excitations than the vacuum ones at the largest $\tilde
k^2$, but {\em much larger} field excitations in the infrared.

\subsection{Yang-Mills theory: Coulomb gauge correlators}

\begin{figure}[p]
\centerline{\epsfxsize=3.0in\epsfbox{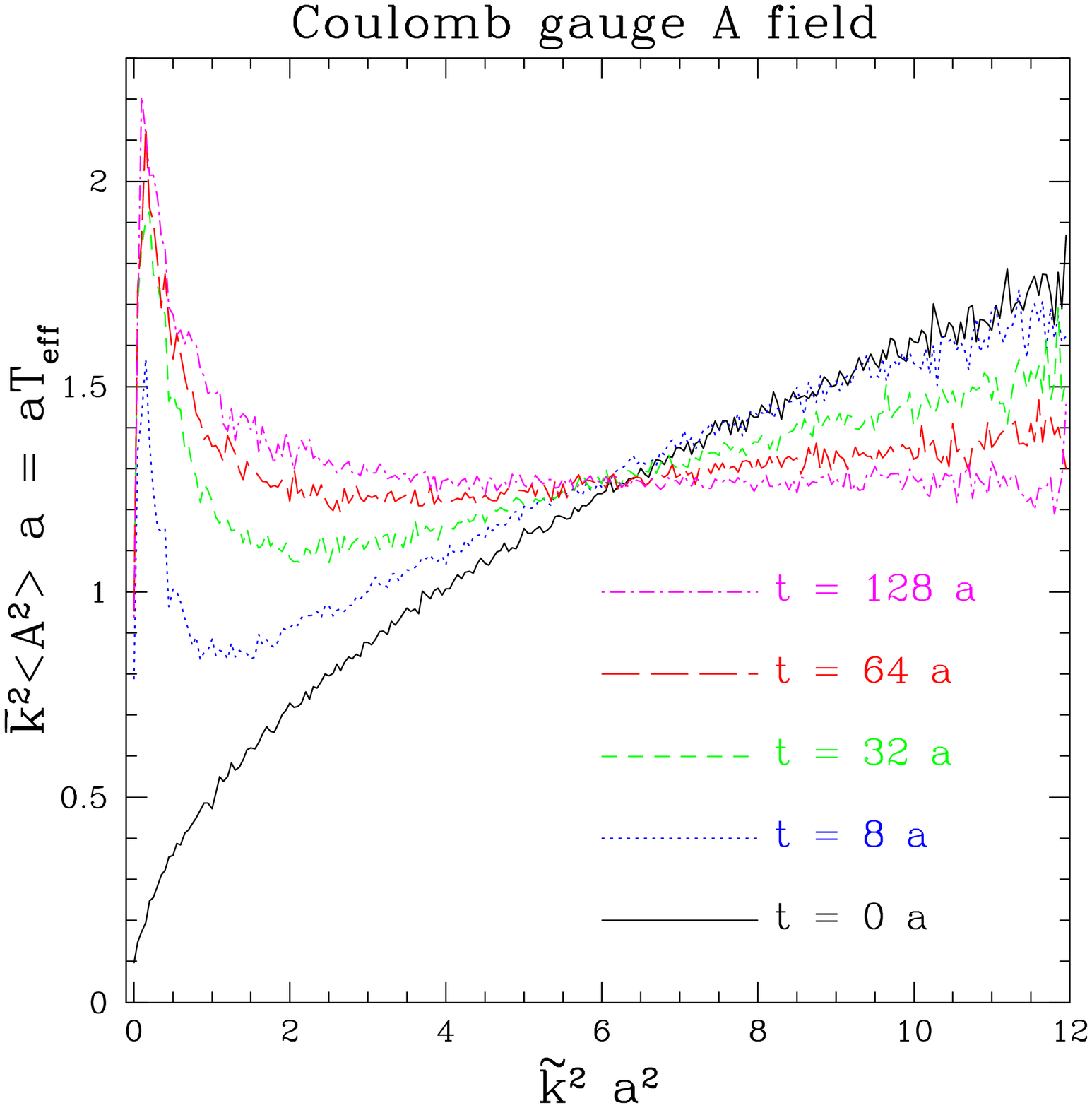} \hspace{0.3in}
\epsfxsize=3.0in\epsfbox{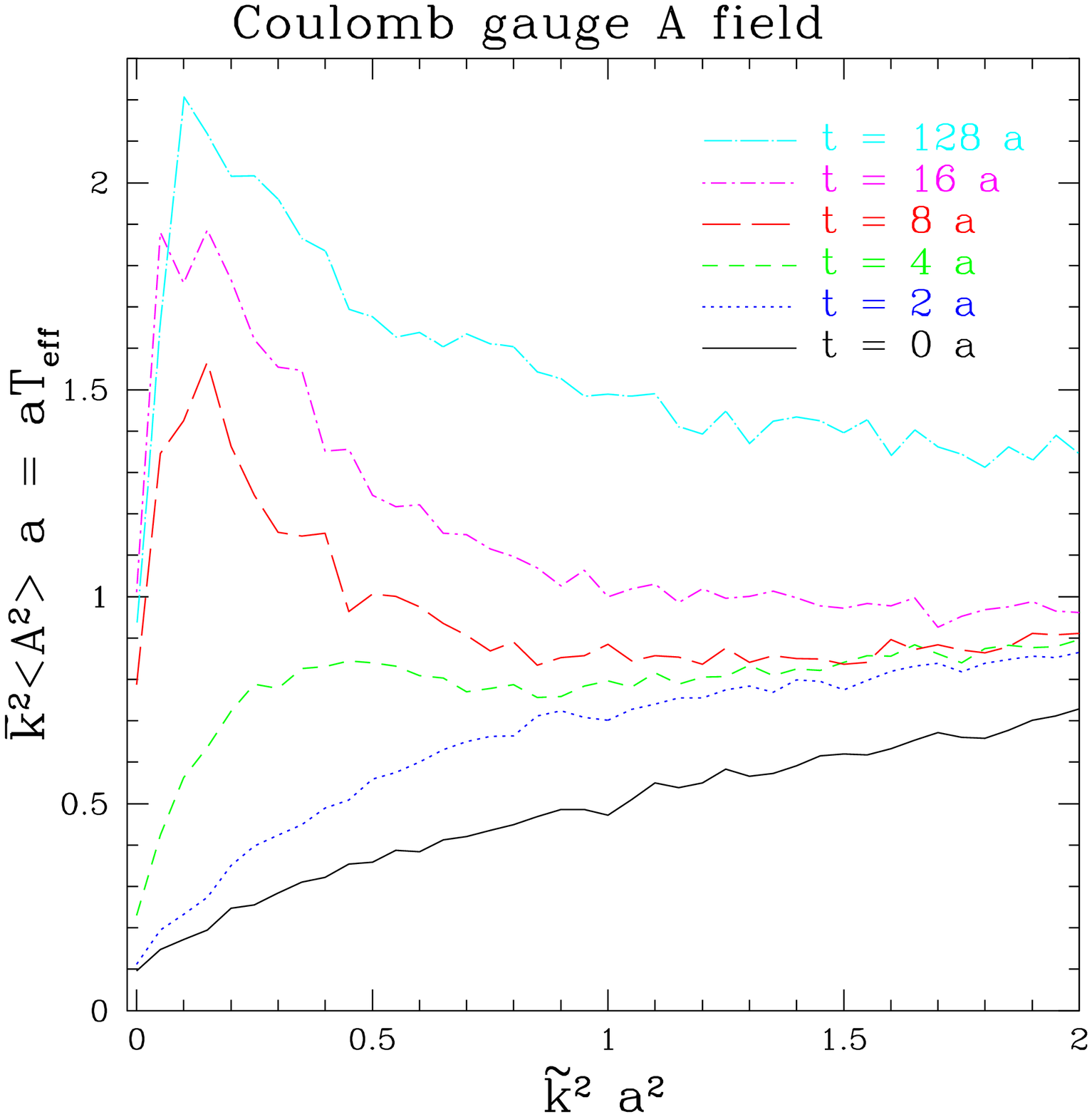}}
\vspace{0.2in}
\centerline{\epsfxsize=3.0in\epsfbox{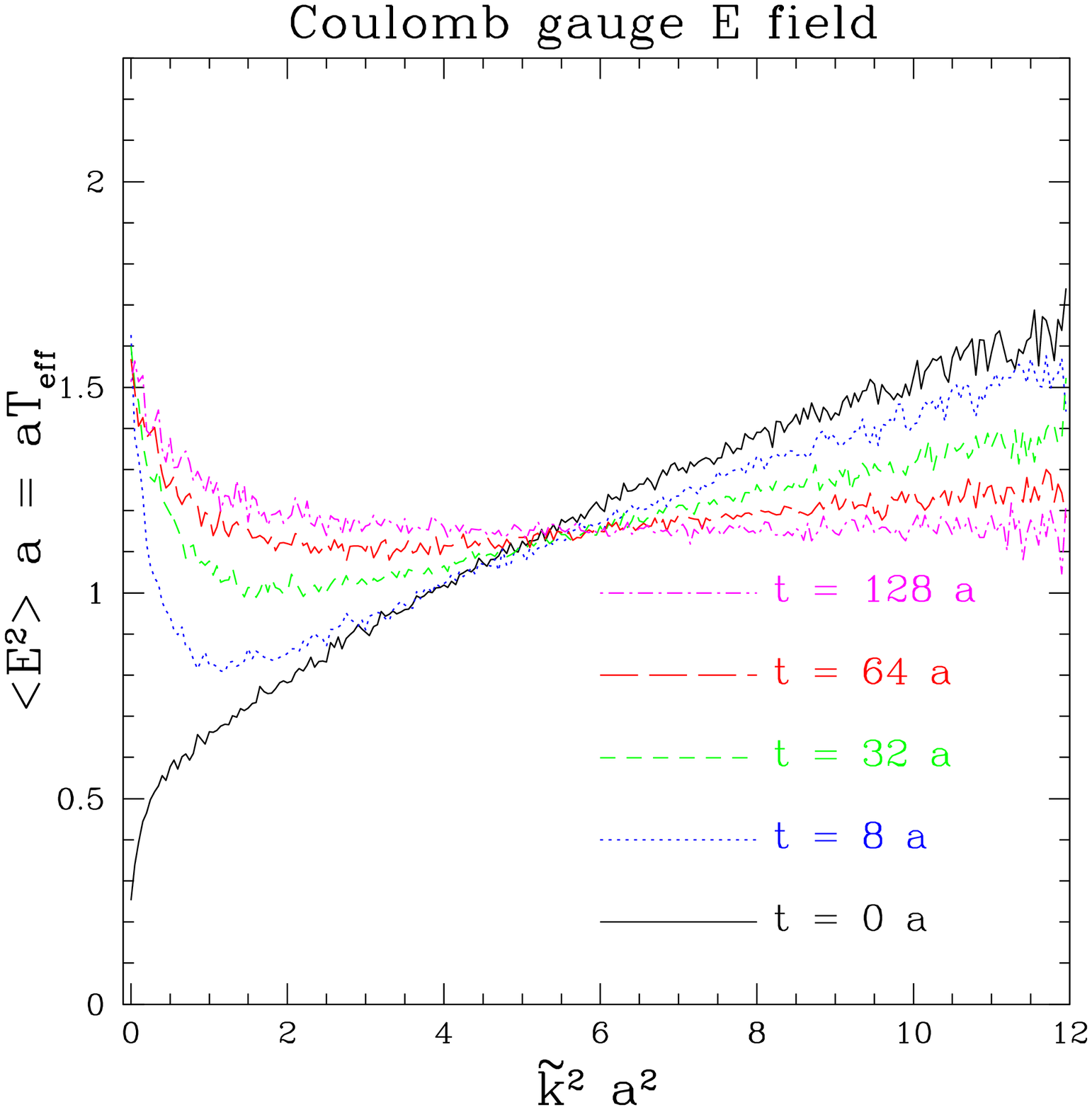} \hspace{0.3in}
\epsfxsize=3.0in\epsfbox{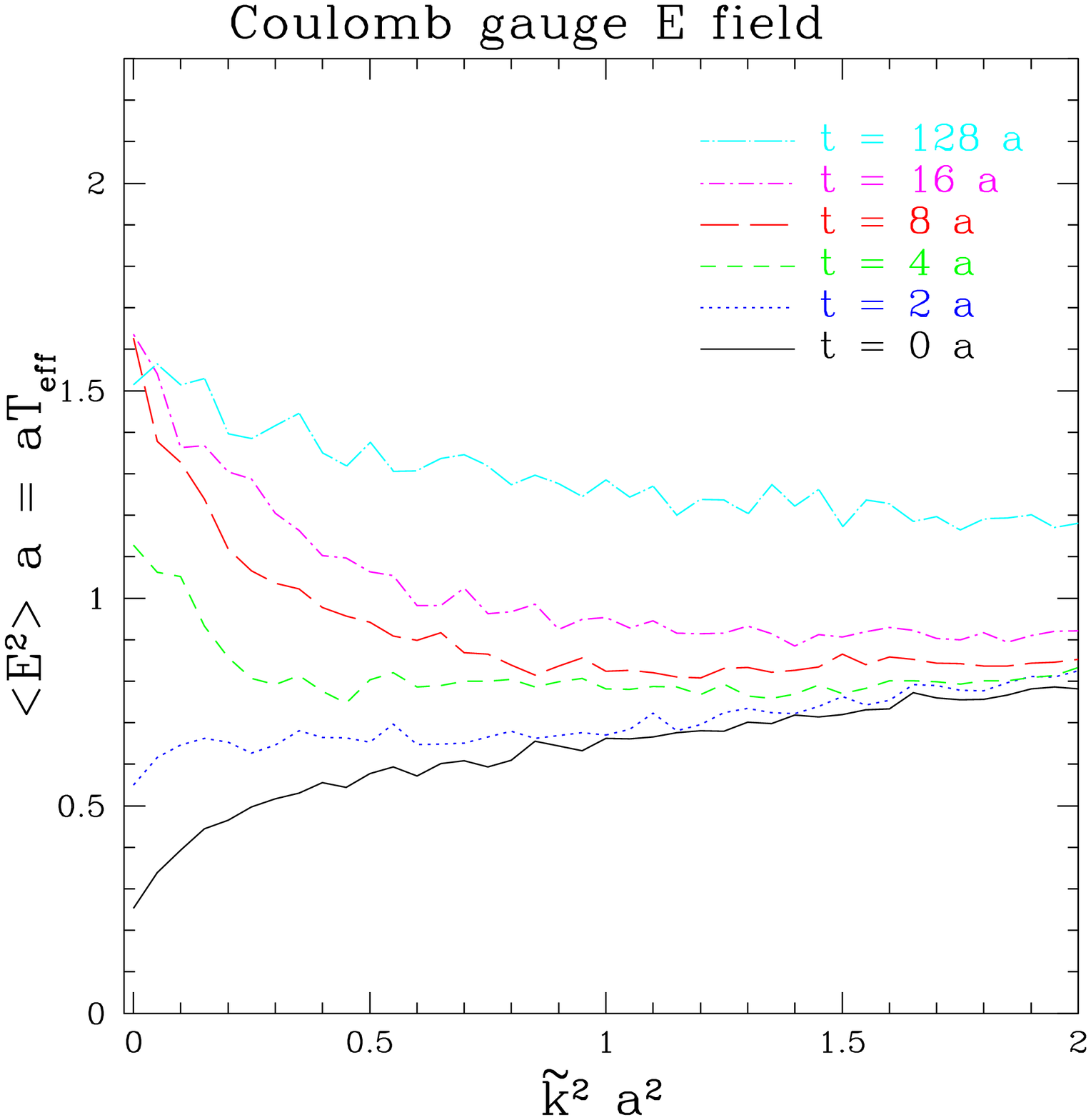}}
\vspace{0.2in}
\caption{Equal time correlators in Fourier space, for
classical lattice SU(2) theory at realistic coupling $g_{\rm w}=0.64$, 
evaluated in Coulomb gauge, at a series of times after ``vacuum'' 
initial conditions are drawn
as described in the text.  Top panels show $\langle \tilde k^2 A^2(k,t)
\rangle$, bottom panels show $\langle E^2(k,t) \rangle$; up to
interaction effects, the results can be interpreted as ``$k$ dependent
temperatures.''  The right
panels show a close up of the small $\tilde{k}^2$ region, and earlier
times.  In equilibrium at zero coupling each curve would be flat; the
equilibrium $E$ field curve is not flat because of Gauss' Law, and the
$A$ field curve is not flat because of interactions.
\label{fig1}}
\end{figure}

This is potentially problematic for the case at hand because the very
infrared physics is the physics of interest.  Eventually, the energy in
the UV modes will cascade into the IR modes in a manner impossible in
the actual quantum theory of interest.  The question is:  On what time
scale does this occur?  We should test whether this takes place
on a time scale short enough to disturb the physics which is of interest
to the simulations.  The theory of
interest is SU(2) Higgs theory in 3+1 dimensions, at realistic coupling.  
A simple way to test the viability of the classical technique 
is to examine the case where there is no condensate, to see how the
``vacuum'' fields evolve classically in isolation.  If the time scale
for poor behavior is shorter than the time scale for ``interesting''
behavior when the dynamics of interest is present, there is a problem.

We need some measurables for our study, which can indicate whether the
cascade of energy into the IR has occurred; I choose to 
study the equal time two-point function in Coulomb gauge, as a
function of $k$.%
\footnote
    {We fix Coulomb gauge with the definition
    of Mandula and Ogilvie \cite{Mandula}; but there are algorithmic
    differences appropriate for the real-time setting.  Our algorithm 
    is discussed in \cite{slavepaper}.  The lattice implementation and 
    classical field evolution algorithm follow 
    Ambj{\o}rn et.~al.~\cite{Ambjorn}.
    }
Coulomb gauge is not expected to give sensible results for
{\em unequal} time correlators because the gauge fixing procedure treats
each time slice separately, except for a single global time dependent
gauge freedom.  However, for equal time correlators it should give
sensible results, although it is unclear exactly how to interpret the
most infrared modes.  For wave numbers where perturbation theory is
useful, we can interpret the values of the Coulomb gauge fixed
correlation functions in terms of a $k$ dependent ``effective
temperature,'' similar to what is done in scalar field studies of
inflation.  We see from Eq.~(\ref{eq:therm}) that the
effective $k$ dependent temperature is
\begin{equation}
T_{\rm eff}(k,A) = \frac{1}{2} (\tilde{k}^2+ m^2) 
	\langle A_i^2(k) \rangle \, ,
\qquad
T_{\rm eff}(k,E) = \frac{1}{2} \langle E_i^2(k) \rangle \, ,
\end{equation}
where the factor of $1/2$ accounts for the two transverse modes summed
over in the index trace; the longitudinal mode of $A$ is fixed to zero
by the Coulomb gauge fixing procedure, and the longitudinal mode of $E$
is set zero, to leading order in perturbation theory, by the Gauss
constraint.  The above procedure also defines $T_{\rm eff}$ even where
perturbation theory does not apply, but it is not clear how to interpret
the result.  Due to interactions, even in
equilibrium $T_{\rm eff}(k)$ will not be constant for all $k$, nor will
it be equal for $A$ and $E$ fields; but wherever the behavior is close
to free field behavior, $T_{\rm eff}(k)$ will be approximately flat and
approximately the same for $A$ and $E$ fields.

Since the cascade of energy, in the gauge sector, should be at
least as fast in the full theory as in just Yang-Mills theory, and
since Yang-Mills theory has fewer parameters, I study just Yang-Mills
theory.  Hence, $m^2 = 0$.  
The initial conditions are chosen as follows. 
All field and momentum variables are
chosen as in Eq.~(\ref{eq:vac_init}), then the $E$ fields are projected
to the Gauss constraint surface; the presence of longitudinal $A$ field
excitations in this choice of initial conditions 
is irrelevant because fixing to Coulomb gauge
will remove them.  This is the same procedure for choosing initial
conditions as was used by Rajantie
et.~al.~\cite{Rajantie}.  
Note that there is only one length scale, $a$, in the problem, and the
gauge coupling is $a$ independent, so all behavior scales as $a$ is
changed; the behavior of a mode of wave number $k$ depends only on
$ak$.

Figure \ref{fig1} shows how the field excitations evolve with
time when we choose initial conditions as in Eq.~(\ref{eq:vac_init}) at
gauge coupling $g_w = 0.64$, on a $64^3$ lattice.  
The striking feature in the figure is that the
infrared excitations ($k^2 a^2 < 1/3$) 
are populated nearly to their thermal value very
early after the evolution begins; they are largely populated by $t = 8a$
and fully populated at $t=16a$.  Then, over a much longer time scale,
the ultraviolet modes thermalize among themselves.  

\subsection{Diffusion of Chern-Simons number}

\begin{figure}[t]
\centerline{\epsfxsize=4in\epsfbox{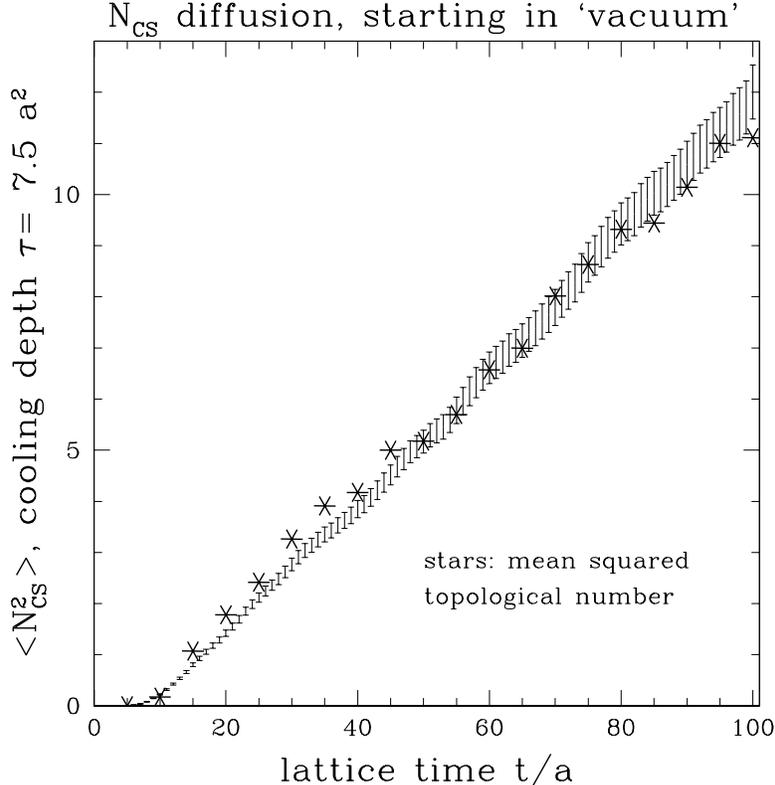}}
\caption{\label{fig2} Diffusion of Chern-Simons number, starting from
``vacuum quantum'' initial conditions, in a $32^3$ box at $g_w=0.64$.
The band of error bars show the mean square value of $\NCS$ as defined in
the text; the stars indicate the mean squared values of the
topological, integer part of $\NCS$.  Efficient baryon number violation
begins at $t \simeq 10 a$.  Data at all $t$ values are from the same set
of time evolutions, so the errors are highly correlated; the integer
measurement was performed on a subset of the data used for the band of
error bars, and have errors about $50\%$ larger.}
\end{figure}

A possible objection to this gauge theory study is that it focused on
gauge fixed observables.
One should always treat gauge fixed measurements with care, so I also
study baryon number violation directly.  The chiral anomaly relates the 
baryon number to the Chern-Simons number,
\begin{equation}
\frac{1}{3} N_{\rm B}(A_i) = \NCS(A_i) \equiv \frac{g^2}{8 \pi^2}
	\int_{A_i'=0}^{A_i'=A_i} \!\!\! d\tau  \int \! d^3 x \;
	\left[ B_i^a(A_\mu ') \, 
	D_\tau {A_i^a} ' \right] \, ,
\label{eq:def_NCS}
\end{equation}
where $\tau$ parameterizes a path through the space of 3-dimensional
gauge fields from the vacuum to the gauge field of interest.  As defined
here $\NCS$ modulo 1 is gauge invariant; the integer part depends on the
path choice.  Choosing the path to coincide with the real time field
evolution, the ensemble average of $\NCS^2$, $\langle \NCS^2 \rangle$,
will grow linearly in time when baryon number violation is active.
Since $\NCS^2$ also has a large ``noise'' contribution from UV
excitations \cite{Ambjorn2}
it is more convenient to measure $\NCS$ not of the actual
configurations in a field evolution but of ``cooled'' copies; this is
discussed in \cite{Ambjorn3}, and extensively in \cite{broken_nonpert},
which also describes the algorithm used here.

Figure \ref{fig2} shows $\langle \NCS^2 \rangle$ as a function of time, 
averaged over about 1000 trajectories with independent
initial conditions, on a $32^3$ lattice and
using a cooling depth (defined in \cite{broken_nonpert}) 
of $7.5 a^2$.  It also
shows the average of the square of the integer part of $\NCS$ (the
stars, which are based on a subset of about 400 of the trajectories,
and therefore have correlated and somewhat larger errors).  The
integer part is defined as the difference between evaluating 
Eq.~(\ref{eq:def_NCS}) for $\NCS(t)$
using the time history as the path, and using the cooling path from
$A_i(t)$ to the 
vacuum as the path.  It is a topological number which directly measures
whether genuine topology change has occurred.  If the vacuum state were
preserved by the time evolution, we would expect $\langle \NCS^2
\rangle$ to fluctuate about an average close to zero, and the
topological number would remain strictly zero.  Instead, 
the figure shows that
there is a brief initial period where this occurs;
then, by $t=10a$, $\NCS$ begins
to diffuse at roughly the same rate it does after full thermalization.
This means that, at the physical value of the coupling $g_w=0.64$, 
the IR gauge fields relevant for baryon number violation
thermalize by about $t = 10 a$, and baryon number proceeds from there.  
This effect is entirely an artifact of using classical lattice evolution
with ``quantum zero point'' initial conditions rather than treating the
actual quantum theory, where baryon number, in vacuum and at this
coupling, is exponentially suppressed with a large exponent. 
The time scale for this ``artifact'' baryon number violation to begin 
quite short; in particular it is {\em shorter} than
the time scale on which Rajantie et.al.~report nontrivial baryon number
violating physics in the Garcia-Bellido scenario \cite{Rajantie}, 
by almost a factor of 3.  This brings seriously into question the
reliability of their results. 

\subsection{Comparison to the scalar theory case}

\begin{figure}[t]
\centerline{\epsfxsize=3.5in\epsfbox{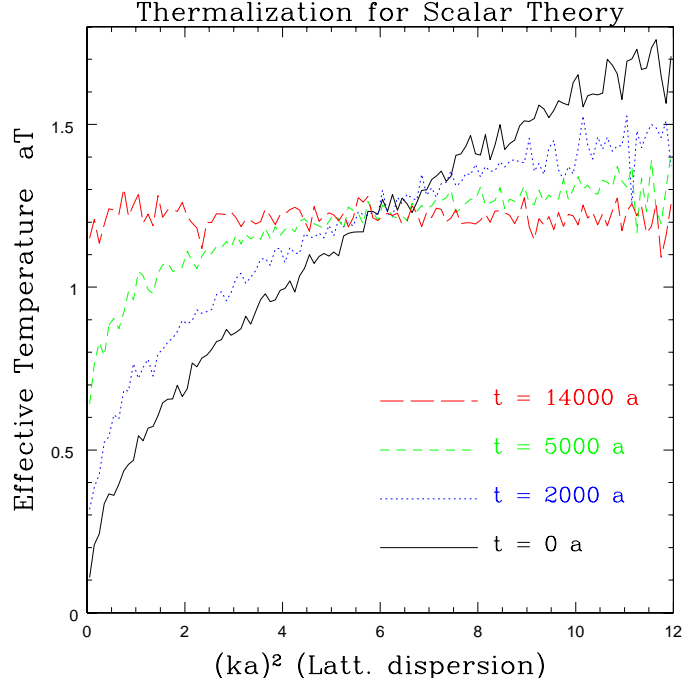}}
\caption{\label{fig3}
Time development of equal time $\dot{\phi}$ correlator in Fourier space
for two component scalar field theory.  Thermalization is slow and does
not proceed first for the infrared modes, as it does in the gauge theory
case.}
\end{figure}

For comparison, and because it bears on the classical field theory
techniques applied in preheating \cite{Khlebnikov,Toms}, I have also
analyzed a two component scalar field theory with Lagrangian  
\begin{equation}
{\cal L} = \int d^3 x \frac{1}{2} \left( \nabla \phi_1^2 + 
	\nabla \phi_2^2 \right) + \frac{m^2}{2}
	\left( \phi_1^2 + \phi_2^2 \right) 
	+ \frac{\lambda}{4} \left( \phi_1^2 + \phi_2^2 \right)^2
	\, .
\end{equation}
This theory behaves very differently from the gauge theory; indeed, even
at quite large coupling the time scale for thermalization turns out to
be very long.  In Figure \ref{fig3} I show how the power spectrum
(which is now a gauge invariant quantity) evolves at the quite large
coupling value of $\lambda = 1$, and bare mass squared
$a^2 m^2 = -.82\lambda$, chosen to
approximately cancel a ``tadpole'' contribution so the initial behavior
is close to that of a massless theory.
The figure shows $T_{\rm eff}(k,P)$, the effective temperature derived
from the momentum degrees of freedom, only; in thermal equilibrium this
will be a flat line (up to statistical fluctuations) 
at any value of the coupling.  The figure shows that, 
in the scalar field case, the thermalization is very slow, even at
$\lambda = 1$, and it proceeds across all $\tilde{k}^2$ values at about
the same rate.  Since the preheating literature generally deals with
much weaker couplings, this suggests that classical lattice techniques
{\em can} be reliable in the preheating context, unless the time scale
of the interesting physics is extremely long.

\section{Discussion}

We have shown that classical Yang-Mills theory, regulated on a lattice
and given ``quantum vacuum'' initial conditions, shows very rapid
heating of the infrared modes, and then more gradual thermalization
between all modes.  The time scale for heating of the infrared, for
SU(2) at realistic coupling $g_{\rm w} = 0.64$, 
is of order 10 lattice spacings of time
(independent of the lattice spacing).  This is probably faster than any
purely infrared, nonequilibrium dynamics of interest will develop, which
makes it difficult to study baryogenesis at preheating without the
results being contaminated by artifacts of the lattice technique.

The physical reason for the rapid transfer of energy between hard and
soft degrees of freedom can best be understood in the language of plasma
physics.  The excitations established for ultraviolet degrees of
freedom, intended to simulate their quantum zero point fluctuations,
propagate nearly ballistically.  However, because the theory is
nonabelian, they carry nonabelian ``charge,'' and constitute a plasma in
which the IR fields evolve.  The plasma degrees of freedom (UV lattice
modes) move in the background of the IR modes of interest, and influence
their evolution rather as a plasma modifies the evolution of infrared
electromagnetic fields.
It has long been appreciated that such a plasma strongly modifies the
dispersion relations of the soft excitations (Debye screening, plasma
oscillations, etc.), 
and efficiently changes the amplitude of soft excitations via
Landau damping.  In the nonabelian context, for quantum,
equilibrium situations, this physics is contained in the ``hard thermal
loops'' of Braaten and Pisarski \cite{BraatenPisarski}.  An
extension to the classical, lattice setting has been addressed by Arnold
\cite{Arnold_latt}.  The damping away of large gauge field excitations
is efficient below the scale $k \sim g/a$ (in a thermal plasma the scale
would be $k \sim gT$).  However, damping is a two way street; the time
scale for a large IR field to be damped away is also the time scale for
UV couplings to populate that IR field, at a ``temperature''
corresponding to some average over the effective temperatures of the UV
degrees of freedom responsible for the damping.  In the simulations of
this paper, no large
initial field exists; so we only see the populating of the IR modes, to
an ``effective temperature'' of order an average over the temperatures
implied by the excitation levels of the UV modes.
This is approximately what we see in Figure \ref{fig1}; the IR modes
with $k \lsim gT$ are
quickly excited to an effective temperature comparable to the effective
temperature of the UV modes; then over a longer time scale, less
infrared modes also thermalize.

Are these lattice artifact physics effects relevant for the study of the
Garcia-Bellido scenario, where besides the ``vacuum'' excitation I have
addressed, there is also large, coherent initial infrared field physics?
I will argue that they are, and that they seriously complicate the
analysis of the Garcia-Bellido proposal by lattice means.  

The interesting physics in the Garcia-Bellido scenario is infrared,
large occupation number physics.  We must make the length scales
associated with that physics (say, the wavelength of modes driven on
resonance) much larger than the lattice spacing, or else the interesting
physics will directly be contaminated by lattice effects.  Therefore the
time scale on which the nontrivial IR dynamics is expected to occur is
quite generically long compared to the lattice spacing $a$.  For
instance, the interesting dynamics in the simulations of Rajantie et.\
al.\ \cite{Rajantie} occurs at $t \sim 30 a$ after the beginning of the
simulation.  Therefore we can expect that the excitation of the IR gauge
field modes by their coupling to the UV should already have taken place
before the interesting IR physics is complete.

If the actual (quantum, nonequilibrium) dynamics of interest fails to
excite large IR gauge fields, then in the simulations we will
nevertheless see large IR gauge fields excited, by the coupling to the
UV described here.  There is therefore clearly the possibility of a
``false positive'' indication of baryon number violation (if for
instance Higgs symmetry is briefly restored, but in the actual dynamics the
gauge fields would be too cold to violate baryon number).  Perhaps less
obviously, it is also possible that the coupling to the UV lattice modes
will create ``false negative'' results where baryon number violation
actually should occur.  This could happen if the true (quantum,
nonequilibrium) dynamics actually excites the IR gauge fields very
efficiently, to an effective temperature higher than what the UV modes
would supply.  In this case, the UV modes represent an efficient
absorber of the IR gauge field excitation energy, via Landau damping.
In the real (vacuum, quantum) theory, energy loss to the UV should not
be too efficient, and should occur mainly by a cascade.  On the lattice,
the UV lattice modes can directly Landau damp away large IR gauge
fields on a time scale $\sim 10 a$, short compared to the (expected)
dynamic timescale for the nontrivial IR dynamics.  Note that the UV
modes also change the dispersion of the lattice modes, which means that
the evolution may be wrong even if energy is not transferred to the UV.
One very basic way of checking for some of these problems is to test for
lattice spacing dependence in the results.  This was not done in
\cite{Rajantie}.

It is possible that there are ways of evading
the problems discussed here.  For instance, one
could choose initial conditions in which only excitations with
$\tilde{k}^2$ below some cutoff were excited.
But it is not obvious that this treatment will reproduce the correct
(quantum) treatment either.  The physics of interest probably involves
energy moving from a condensate into IR modes, and then cascading into
the UV.  Will this cascade be incorrectly
described if the UV modes start out with no excitation?  Does failing to
quantize the UV modes already ensure that the description cannot be
correct?  The burden of proof
clearly lies with the practitioner.  It would be necessary
but not sufficient to demonstrate that all simulation results show weak
dependence on the choice of $\tilde{k}^2$ cutoff.

The behavior we have seen should {\em not} be expected in 1+1
dimensional, abelian studies of baryogenesis at preheating, such as
those of Garcia-Bellido et.~al.~\cite{GB,GB2}.
This is a matter of dimensionality.  In 3+1 dimensions, thermal
self-energy corrections are UV divergent in a classical theory, with the
divergence cut off in nature at $k \sim T$ by quantum mechanics.
However, in 1+1 dimensions thermal self-energies are UV
finite.  Therefore the studies of the 1+1 dimensional abelian analog
theory in \cite{GB,GB2} are deceptive.
The behavior I describe also appears to play little role in abelian
Higgs theory, at least 
at the unrealistically small coupling considered in \cite{Rajantie2}.
I should also emphasize that my results do {\em not} mean that
previous studies of inflationary preheating are incorrect.  As Figure
\ref{fig3} shows, the time scale for UV energy to cascade to the IR in a
scalar theory, even at the very large coupling of $\lambda = 1$, is very
long, thousands of lattice units of time.  
The cascade time is expected to grow at weak coupling as
$\lambda^{-2}$.  The largest value of $\lambda$ used in \cite{Toms} was
about two orders of magnitude smaller than that in Fig.~\ref{fig3}, so
the time scale for the cascade to occur in their work 
would be at least $ 10^7$ lattice units, much longer than any time scale
they considered.
Physically this difference is related to the fact that the hard thermal
loop for a gauge field contains an imaginary (Landau damping) part,
while that for a scalar field does not.

\section*{acknowledgments}

I thank Arttu Rajantie for discussions.  
This work was supported, in part, by the U.S. Department
of Energy under Grant Nos.~DE-FG03-96ER40956.

\end{document}